\documentclass[11pt,reqno,oneside]{amsart}
\usepackage{verbatim,amsmath,amssymb,cite,epsfig}
\title{Stability of Evolutionarily Stable Strategies in Discrete Replicator Dynamics
with Time Delay} 
\author{Jan Alboszta and Jacek Mi\c{e}kisz \\ Institute of Applied Mathematics \\
and Mechanics \\ Warsaw University  \\ ul. Banacha 2  \\ 02-097
Warsaw, Poland} 
\begin{document} 
\baselineskip=20pt
\maketitle 

\noindent {\bf Abstract}: We construct two models of discrete-time replicator dynamics
with time delay. In the social-type model, players imitate opponents 
taking into account average payoffs of games played some units of time ago.
In the biological-type model, new players are born from parents who played
in the past. We consider two-player games with two strategies and a unique mixed 
evolutionarily stable strategy. We show that in the first type of dynamics,
it is asymptotically stable for small time delays and becomes unstable for big ones
when the population oscillates around its stationary state. 
In the second type of dynamics, however, evolutionarily stable strategy 
is asymptotically stable for any size of a time delay. 
\vspace{3mm}

\vspace{3mm}

\noindent Corresponding author: Jacek Mi\c{e}kisz, e-mail: miekisz@mimuw.edu.pl, 
phone: 48-22-5544423, fax: 48-22-5544300.
\eject

\newtheorem{theo}{Theorem}
\newtheorem{defi}{Definition}
\newtheorem{hypo}{Hypothesis}

\section{Introduction}
\numberwithin{equation}{section}

\noindent The evolution of populations can be often described within game-theoretic models.
The pioneer of such approach, John Maynard Smith, introduced the fundamental notion 
of an evolutionarily stable strategy (1974, 1982). If everybody plays such a strategy, 
then the small number of mutants playing a different strategy is eliminated 
from the population. The dynamical interpretation of the evolutionarily stable strategy
was later provided in (Taylor and Jonker, 1978; Hofbauer {\em et al.}, 1979;
Zeeman, 1981). A system of differential or difference equations, 
the so-called replicator equations, were proposed. They describe the time-evolution 
of frequencies of strategies. It is known that any evolutionarily stable strategy 
is an asymptotically stable stationary point of such dynamics (Hofbauer and Sigmund, 1988); 
Weibull, 1997). In fact, in two-player games with two strategies, a state of population 
is evolutionarily stable if and only if it is asymptotically stable. 

Recently Tao and Wang (1997) investigated the effect of a time delay 
on the stability of interior stationary points of the replicator dynamics. 
They considered a two-player game with two strategies and a unique asymptotically 
stable interior stationary point. They proposed a certain form of a time-delay 
differential replicator equation. They showed that the mixed evolutionarily stable strategy 
is stable if a time delay is small. For sufficiently large delays it becomes unstable. 

Here we discuss discrete-time replicator dynamics. We introduce a time delay 
in two different ways. In the so-called social model, we assume that players 
imitate a strategy with the higher average payoff, taking into account time-delayed information. 
When the limit of zero time interval is taken, one obtains equations 
discussed by Tao and Wang (1997). In the so-called biological model, 
we assume that the number of players born in a given time is proportional 
to payoffs received by their parents in a certain moment in the past. 
We analyze the stability of the interior stationary point in our discrete time-delay 
replicator dynamics. In the first model, we prove results analogous 
to those of Tao and Wang (1997). However, in our second model, we show the stability 
of the interior point for any value of the time delay. We prove our results 
in an elementary way, that is without any reference to the theory of time-delay equations.

In Section 2, we present our discrete-time replicator dynamics with two types of time delay.
In Section 3, we discuss a model with a social-type time delay and in Section 4, 
with a biological-type time delay. Discussion follows in Section 5. Proofs are given
in Appendix A and B.
    
\section{Replicator dynamics}

\noindent We assume that our population is haploid, that is the offspring  
have identical phenotypic strategies as their parents 
(we assume that there are two different pure strategies; denote them by $A$ and $B$).
In discrete moments of time, individuals compete in pairwise contests 
and the outcome is given by the following payoff matrix:  
\vspace{3mm}

\hspace{23mm} A  \hspace{2mm} B   

\hspace{15mm} A \hspace{3mm} a  \hspace{3mm} b 

U = \hspace{6mm} 

\hspace{15mm} B \hspace{3mm} c  \hspace{3mm} d,

where the $ij$ entry, $i,j = A, B$, is the payoff of the first (row) player when
he plays the strategy $i$ and the second (column) player plays the strategy $j$. 
We assume that both players are the same and hence payoffs of the column player are given 
by the matrix transposed to $U$; such games are called symmetric. From now on
we will assume that $c>a$ and $b>d$ so there exists a unique 
mixed evolutionarily stable strategy, $x^{*}=\frac{\delta_{2}}{\delta}$, 
where $\delta_{1}=c-a, \delta_{2}=b-d$, and $\delta=\delta_{1}+\delta_{2}$. 
$x^{*}$ is the equilibrium fraction of an infinite population playing $A$. 

We will first discuss replicator dynamics without a time delay.
We assume that during the small time interval $\epsilon$, only an $\epsilon$ fraction 
of the population takes part in pairwise competitions, that is plays games.
Let $p_{i}(t)$, $i=A, B,$ be the number of individuals playing at the time $t$ the strategy 
$A$ and $B$ respectively, $p(t)=p_{A}(t)+p_{B}(t)$ the total number of players and
$x(t)=\frac{p_{1}(t)}{p(t)}$ a fraction of the population playing $A$.
We write
\begin{equation}
p_{i}(t + \epsilon) = (1-\epsilon)p_{i}(t) + \epsilon p_{i}(t)U_{i}(t); \; \; i= A,B,
\end{equation}

where $U_{A}(t)= ax(t)+b(1-x(t))$ and $U_{B}(t)= cx(t)+d(1-x(t))$  
are average payoffs of individuals playing A and B respectively.
We assume that all payoffs are not smaller than $0$ hence $p_{A}$ and $p_{B}$
are always nonnegative and therefore $0\leq x \leq 1$. 

The equation for the total number of players reads

\begin{equation}
p(t + \epsilon) = (1-\epsilon)p(t) + \epsilon p(t)\bar{U}(t),
\end{equation}
where $\bar{U}(t)=x(t)U_{A}(t)+(1-x(t))U_{B}(t)$ is the average payoff in the population.
When we divide (2.1) by (2.2) we obtain an equation for the frequency of the strategy $A$,

\begin{equation}
x(t + \epsilon) - x(t) = \epsilon\frac{x(t)[U_{A}(t) - \bar{U}(t)]}
{1-\epsilon + \epsilon \bar{U}(t)}.
\end{equation}

Now we divide both sides of (2.3) by $\epsilon$, perform the limit $\epsilon \rightarrow 0$,
and obtain the well known differential replicator equation,

\begin{equation}
\frac{dx(t)}{dt}=x(t)[U_{A}(t) - \bar{U}(t)].
\end{equation}

The above equation can also be written as
\begin{equation}
\frac{dx(t)}{dt}=x(t)(1-x(t))[U_{A}(t) - U_{B}(t)]= -\delta x(t)(1-x(t))(x(t)-x^{*}).
\end{equation}

It follows that $x^{*}$ is the globally (except the boundary points $x=0$ and $x=1$) asymptotically stable
stationary point of (2.5). 

In the following two sections we will study the stability of $x^{*}$ in the replicator dynamics
with different forms of the time delay.

\section{Social-type time delay}

\noindent Here we assume that individuals at time $t$ replicate due to average payoffs 
obtained by their strategies at time $t-\tau$ for some delay $\tau>0$
(see also a discussion after (3.6)). 

We propose the following equations:

\begin{equation}
p_{i}(t + \epsilon) = (1-\epsilon)p_{i}(t) + \epsilon p_{i}(t)U_{i}(t-\tau); \; \; i= A,B.
\end{equation}
Then for the total number of players we get
\begin{equation}
p(t + \epsilon) = (1-\epsilon)p(t) + \epsilon p(t)\bar{U}_{o}(t-\tau),
\end{equation}
where $\bar{U}_{o}(t-\tau)=x(t)U_{A}(t-\tau)+(1-x(t))U_{B}(t-\tau).$

We divide (3.1) by (3.2) and obtain an equation for the frequency of the strategy $A$,

\begin{equation}
x(t + \epsilon) - x(t) = \epsilon \frac{x(t)[U_{A}(t-\tau) - \bar{U}_{o}(t-\tau)]}
{1-\epsilon + \epsilon \bar{U}_{o}(t-\tau)} 
\end{equation}

and after some rearrangements we get

\begin{equation}
x(t + \epsilon) - x(t) = -\epsilon x(t)(1-x(t))[x(t-\tau)-x^{*}]\frac{\delta}{1-\epsilon + 
\epsilon \bar{U}_{o}(t-\tau)}. 
\end{equation}

Now the corresponding replicator dynamics in the continuous time reads

\begin{equation}
\frac{dx(t)}{dt}=x(t)[U_{A}(t-\tau) - \bar{U}_{o}(t-\tau)]
\end{equation}

and can also be written as

\begin{equation}
\frac{dx(t)}{dt}=x(t)(1-x(t))[U_{A}(t-\tau) - U_{B}(t-\tau)]=-\delta x(t)(1-x(t))(x(t-\tau)-x^{*}).
\end{equation}

The first equation in (3.6) can be also interpreted as follows. Assume that randomly chosen players 
imitate randomly chosen opponents. Then the probability that a player who played $A$ 
would imitate the opponent who played $B$ at time $t$ is exactly $x(t)(1-x(t)).$ 
The intensity of imitation depends on the delayed information about the difference 
of corresponding payoffs at time $t- \tau$. We will therefore say that such models 
have a social-type time delay.

Equations (3.5-3.6) are exactly the time-delay replicator dynamics proposed and analyzed 
by Tao and Wang (1997). They showed that if 
$\tau< \delta \pi /2 \delta_{1} \delta_{2}$,
then the mixed evolutionarily stable strategy, $x^{*}$, is asymptotically stable. 
When $\tau$ increases beyond the bifurcation value $\delta \pi /2 \delta_{1}\delta_{2}$, $x^{*}$ 
becomes unstable. We will show in Appendix A analogous results for our discrete equation (3.4). 
Our proofs will be elementary - we will not refer to the theory of time-delay equations.
In particular, we will show that for any large enough time delay, $\tau$, $x(t)$ 
oscillates around $x^{*}$.
 
\section{Biological-type time delay}

\noindent Here we assume that individuals born at time $t-\tau$ are able to take part in contests 
when they become mature at time $t$ or equivalently they are born $\tau$ units of time 
after their parents played and received payoffs. We propose the following equations:

\begin{equation}
p_{i}(t + \epsilon) = (1-\epsilon)p_{i}(t) + \epsilon p_{i}(t-\tau)U_{i}(t-\tau); \; \; i= A,B.
\end{equation}
Then the equation for the total number of players reads

\begin{equation}
p(t + \epsilon) = (1-\epsilon)p(t) + \epsilon p(t)[\frac{x(t)p_{A}(t-\tau)}{p_{A}(t)}U_{A}(t-\tau)
+\frac{(1-x(t))p_{B}(t-\tau)}{p_{B}(t)}U_{B}(t-\tau)].
\end{equation}

\noindent We divide (4.1) by (4.2) and obtain an equation for the frequency of the first strategy,

\begin{equation}
x(t + \epsilon) - x(t) = \epsilon \frac{x(t - \tau)U_{A}(t - \tau) - x(t)\bar{U}(t - \tau)}
{(1-\epsilon)\frac{p(t)}{p(t-\tau)} + \epsilon \bar{U}(t-\tau)},
\end{equation}

\noindent where $\bar{U}(t-\tau)=x(t-\tau)U_{A}(t-\tau)+(1-x(t-\tau))U_{B}(t-\tau).$
We will prove in Appendix B that $x^{*}$ is asymptotically stable for any value 
of the time delay $\tau$. Here we will show our result in the following simple example.
\vspace{3mm}

\noindent The payoff matrix is given by  
$U = \left(\begin{array}{cc}
    0 & 1\\
    1 & 0
  \end{array}\right)$ hence $x^{*} = \frac{1}{2}$ is the mixed evolutionarily stable strategy
which is asymptotically stable in the replicator dynamics without time delay.
The equation (4.3) now reads

\begin{equation}
x(t+\epsilon)-x(t)=\epsilon \frac{x(t-\tau)(1 - x(t-\tau))-2 x(t)x(t-\tau)(1-x(t-\tau))}
{(1-\epsilon)\frac{p(t)}{p(t-\tau)}+2\epsilon x(t-\tau)(1-x(t-\tau))}. 
\end{equation}

After simple algebra we get

\begin{equation}
x(t+ \epsilon) - \frac{1}{2} + \frac{1}{2} - x(t) =
 \epsilon (1 - 2x(t)) \frac{x(t-\tau)(1-x(t - \tau))}
{(1-\epsilon)\frac{p(t)}{p(t - \tau)} + 2\epsilon x(t - \tau)(1 - x (t - \tau))},
\end{equation}
$$x(t + \epsilon) - \frac{1}{2} = (x(t)- \frac{1}{2}) \frac{1}
{1 + \frac{\epsilon p(t - \tau)}{(1-\epsilon)p(t)}2x(t - \tau)(1- x(t - \tau))}$$ 

hence

\begin{equation}
|x(t+\epsilon)-\frac{1}{2}| < |x(t)-\frac{1}{2}|. 
\end{equation}
  
It follows that $x^{*}$ is globally asymptotically stable.

\section{Discussion}

\noindent We proposed and analyzed certain forms of discrete-time replicator dynamics
with a time delay. We introduced the time delay on the level of the number of individuals 
playing different strategies and not on the level of their relative frequencies.
We showed that the stability of a mixed evolutionarily stable strategy depends
on the particular form of the time delay.   

It would be interesting to analyze time-delay replicator models taking into account
the limited capacity of an environment (for example in the form of a logistic equation)
and stochastic effects resulting from mutations and a random character of interactions.
In the latter case, individuals play with particular opponents and not against 
the average strategy like in the standard replicator dynamics.  
\vspace{3mm}

\noindent {\bf Acknowledgments} JM would like to thank 
the Polish Committee for Scientific Research for a financial support
under the grant KBN 5 P03A 025 20.

\appendix 
\section{} 

\begin{theo}
$x^{*}$ is asymptotically stable in the dynamics (3.4) if $\tau$ is sufficiently small
and unstable for large enough $\tau$.
\end{theo}

\noindent {\bf Proof:} We will assume that $\tau$ is a multiple of $\epsilon$, 
$\tau=m\epsilon$ for some natural number $m$. Observe first that if $x(t - \tau) < x^{*}$, 
then $x(t + \epsilon) > x(t)$, and if $x(t - \tau) > x^{*}$, 
then $x(t + \epsilon) < x(t)$. Let us assume first that there is $t'$ 
such that $x(t'), x(t'-\epsilon), x(t'-2\epsilon),..., x(t'-\tau) < x^{*}$.
Then $x(t)$, $t \geq t'$ increases up to the moment $t_{1}$ for which
$x(t_{1} - \tau) > x^{*}$. If such $t_{1}$ does not exist then 
$x(t) \rightarrow_{t \rightarrow \infty}x^{*}$ and the theorem is proved.
Now we have $x^{*}<x(t_{1}-\tau) < x(t_{1}-\tau+\epsilon)< \ldots <x(t_{1})$
and $x(t_{1}+\epsilon)<x(t_{1})$ so $t_{1}$ is a turning point. Now $x(t)$
decreases up to the moment $t_{2}$ for which $x(t_{2}-\tau) < x^{*}$.
Again, if such $t_{2}$ does not exist, then the theorem follows.
Therefore let us assume that there is an infinite sequence, $t_{i}$, of such turning points.
Let $\eta_i = |x(t_{i}) - x^{*}|$. We will show that 
$\eta_i \rightarrow_{i \rightarrow \infty} 0$.
  
For $t \in \{t_{i}, t_{i} + \epsilon, \ldots, t_{i + 1}-1\}$ we have the following bound 
for $x(t + \epsilon) - x(t)$:
 
\begin{equation}
|x(t + \epsilon) - x(t)| < \frac{1}{4} \eta_{i} \frac{\epsilon \delta}
{1 -\epsilon + \epsilon \bar{U}_{o}(t - \tau)}.
\end{equation}
  
This means that

\begin{equation}
\eta_{i+1} <  (m+1)\epsilon K \eta_{i},
\end{equation}  

where $K$ is the maximal possible value of 
$\frac{\delta}{4(1-\epsilon + \epsilon \bar{U}_{o}(t - \tau))}.$
We get that if 

\begin{equation}
\tau < \frac{1}{K}-\epsilon,
\end{equation}    
then $\eta_{i}\rightarrow_{i \rightarrow \infty} 0$ so $x(t)$ converges to $x^{*}$. 

Now if for every $t$, 
$|x(t + \epsilon)-x^{*}| < \max_{k\in \{0,1,...,m\}}|x(t - k\epsilon)-x^{*}|$, 
then $x(t)$ converges to $x^{*}$. 
Therefore assume that there is $t''$ such that
$|x(t'' + \epsilon)-x^{*}| \geq \max_{k\in \{0,1,...,m\}}|x(t'' - k\epsilon)-x^{*}|$. 
If $\tau$ satisfies (A.3), then it follows that 
$x(t+\epsilon),...,x(t+\epsilon+\tau)$ are all on the same side of $x^{*}$ 
and the first part of the proof can be applied. We showed that $x(t)$ converges to $x^{*}$ 
for any initial conditions different from $0$ and $1$ hence 
$x^{*}$ is globally asymptotically stable.

Now we will show that $x^{*}$ is unstable for any large enough $\tau$.

Let $\gamma>0$ be arbitrarily small and consider a following perturbation 
of the stationary point $x^{*}$: $x(t)=x^{*}, t\leq 0$ 
and $x(\epsilon) = x^{*}+\gamma$. It folows from (3.4) 
that $x(k\epsilon)=x(\epsilon)$ for $k=1,...,m+1$. 
Let $K'=\min_{x \in[x^{*}-\gamma,x^{*}+\gamma]}
\frac{x(1-x)\delta}{4(1-\epsilon + \epsilon \bar{U}_{o}(t - \tau))}$. 
If $\frac{m}{2}\epsilon K'\gamma > 2\gamma$, that is 
$\tau > \frac{4}{K'}$, then it follows from (3.4) 
that after $m/2$ steps (we assume without loss of generality that $m$ is even)
$x((m+1+m/2)\epsilon)<x^{*}-\gamma$. In fact we have 
$x((2m+1)\epsilon)< \ldots < x((m+1)\epsilon)$ and at least $m/2$ of $x's$ in this sequence
are smaller than $x^{*}-\gamma$. Let $\bar{t}>(2m+1)\epsilon$ be the smallest 
$t$ such that $x(t)>x^{*}-\gamma$. Then we have 
$x(\bar{t}-m \epsilon), \ldots, x(\bar{t}-\epsilon) < x^{*}-\gamma < x(\bar{t})$
hence after $m/2$ steps, $x(t)$ crosses $x^{*}+\gamma$ and the situation repeats itself.
We showed that if
\begin{equation}
\tau >\frac{4}{K'}, 
\end{equation}   
then there exists an infinite sequence, $\tilde{t}_{i}$, such that $|x(\tilde{t}_{i})-x^{*}|>\gamma$
and therefore $x^{*}$ is unstable. Moreover, $x(t)$ oscillates around $x^{*}$.

\section{}

\begin{theo}
$x^{*}$ is asymptotically stable for any $\tau$ in the dynamics (4.3).  
\end{theo}

\noindent {\bf Proof:} Let $c_{t} = \frac{x(t)U_{A}(t)}{\bar{U}(t)}$. 
Observe that if $x(t) < x^{*}$, then $c_{t} > x(t)$, if $x(t) > x^{*}$, 
then $c_{t} < x(t)$, and if $x(t) = x^{*}$, then $c_{t} = x^{*}$. 
We can write (4.3) as

\begin{equation}
x(t + \epsilon) - x(t) = \epsilon\frac{x(t - \tau)U_{A}(t - \tau) - c_{t - \tau}
\bar{U}(t - \tau) + c_{t - \tau} \bar{U}(t - \tau) - x(t) \bar{U}(t - \tau)}
{(1-\epsilon)\frac{p(t)}{p(t - \tau)} + \epsilon \bar{U}(t - \tau)}
\end{equation}
and after some rearrangements we obtain

\begin{equation}
x(t + \epsilon) - c_{t - \tau} = (x(t) - c_{t - \tau}) \frac{1}
{1 + \frac{\epsilon p(t - \tau)}{(1-\epsilon) p(t)} \bar{U}(t - \tau)}. 
\end{equation}

We get that at time $t + \epsilon$, $x$ is closer to $c_{t -\tau}$ than at time $t$
and it is on the same side of $c_{t -\tau}$.
We will show that  $c$ is an increasing or a constant function of $x$. 
Let us calculate the derivative of $c$ with respect to $x$. 

\begin{equation}
c' = \frac{f(x)}{(xU_{A} + (1 - x)U_{B})^2}, 
\end{equation}
where  
\begin{equation}
f(x)= (ac + bd - 2ad)x^{2}+ 2d(a-b)x+bd.
\end{equation}

A simple analysis shows that $f>0$ on $(0,1)$ or $f=0$ on $(0,1)$
(in the case of $a=d=0$). Hence $c(x)$ is either an increasing
or a constant function of $x$. In the latter case, $\forall_x c(x) = x^{*},$ 
as it happens in our example in Section 4, and the theorem follows.
  
We will now show that
\begin{equation}
|x(t + \tau + \epsilon) - x^{*}| < \max\{|x(t) - x^{*}|, |x(t + \tau) - x^{*}|\} 
\end{equation}
hence $x(t)$ converges to $x^{*}$ for any initial conditions different from
$0$ and $1$ so $x^{*}$ is globally asymptotically stable.

If $x(t) < x^{*}$ and $x(t + \tau) <  x^{*}$, then $x(t) < c_{t} \leq x^{*}$ 
and also $x(t+\tau) < c_{t+\tau} \leq x^{*}$. From (B.2) we obtain

  \[ \left\{\begin{array}{cc}
       x \left( t + \tau \right) < x \left( t + \tau + \epsilon \right) < c_t 
       \; \; if \; \;  x \left( t + \tau \right) < c_t\\
       x \left( t \right) < x \left( t + \tau + \epsilon \right) = c_t  \; \; if \; \; 
       x \left( t + \tau \right) = c_t\\
       x \left( t \right) < c_t < x \left( t + \tau + \epsilon \right) < x \left( t +
       \tau \right)\; \;  if \; \;  x \left( t + \tau \right) > c_t
     \end{array}\right. \]
hence (B.5) holds.

If $x \left( t \right) > x^{\ast}$ and $x \left( t + \tau \right) <
  x^{\ast}$, then $x \left( t + \tau \right) < x^{\ast} < c_t < x \left( t
  \right)$ and either $x \left( t + \tau \right) <
  x \left( t + \tau + \epsilon \right) < x^{\ast}$ or $x^{\ast} < x \left( t + \tau
  + \epsilon \right) < c_t$ which means that (B.5) holds.
  
The cases of $x \left( t \right) > x^{*}$, $x \left( t + \tau \right) >
  x^{*}$ and $x \left( t \right) < x^{*}$, $x \left( t + \tau \right)<x^{*}$
can be treated analogously. We showed that (B.5) holds.
\vspace{3mm}

\noindent {\bf References}
\vspace{3mm}

\noindent Hofbauer, J., Shuster, P. Sigmund, K., 1979.
A note on evolutionarily stable strategies and game dynamics.
J. Theor. Biol. 81, 609-612.
\vspace{2mm}

\noindent Hofbauer, J., Sigmund, K., 1988. The Theory
of Evolution and Dynamical Systems. Cambridge University Press, Cambridge.
\vspace{2mm}
 
\noindent Maynard Smith, J., 1974. The theory of games and the evolution
of animal conflicts. J. Theor. Biol. 47, 209-221.
\vspace{2mm}

\noindent Maynard Smith, J., 1982. Evolution and the Theory of Games.
Cambridge University Press, Cambridge.
\vspace{2mm}

\noindent Tao, Y. and Wang, Z., 1997.
Effect of time delay and evolutionarily stable strategy.
J. Theor. Biol. 187, 111-116.
\vspace{2mm}

\noindent Taylor, P. D., Jonker, L. B., 1978. Evolutionarily stable
strategy and game dynamics. Math. Biosci. 40, 145-156.
\vspace{2mm}

\noindent Weibull, J., 1997. Evolutionary Game Theory. 
MIT Press, Cambridge.
\vspace{2mm}

\noindent Zeeman, E., 1981. Dynamics of the evolution of animal conflicts.
J. Theor. Biol. 89, 249-270.

\end{document}